\documentclass[conference]{IEEEtran}

\usepackage{algorithm}
\usepackage{algpseudocode}
\usepackage{slashbox}
\usepackage{amssymb}
\usepackage{algpseudocode}
\usepackage{subfloat}
\usepackage{subfig}
\usepackage{amsmath}
\usepackage{array}
\usepackage[english]{babel}
\usepackage{mdwmath}
\usepackage{algpseudocode}
\usepackage{enumerate}
\usepackage{booktabs}
\usepackage{color}
\usepackage{amsthm}
\theoremstyle{definition}

\theoremstyle{remark}

\usepackage{amssymb}
\usepackage{lineno}
\usepackage{array}
\newcolumntype{L}[1]{>{\raggedright\let\newline\\\arraybackslash\hspace{0pt}}m{#1}}
\newcolumntype{C}[1]{>{\centering\let\newline\\\arraybackslash\hspace{0pt}}m{#1}}
\newcolumntype{R}[1]{>{\raggedleft\let\newline\\\arraybackslash\hspace{0pt}}m{#1}}
\usepackage[utf8]{inputenc}
\usepackage{paralist}
\usepackage[usenames,dvipsnames]{xcolor}
\usepackage{epstopdf}
\usepackage{lipsum}% http://ctan.org/pkg/lipsum
\usepackage{multicol}% http://ctan.org/pkg/multicol
\usepackage[figuresright]{rotating}
\usepackage[textsize=footnotesize]{todonotes}

% ...

\begin{document}
%

%\begin{equation}
%  \left\lbrace \rule{3cm}{1cm}\right.
%\end{equation}
%
%\begin{equation}
%  \left. \rule{3cm}{1cm}\right\rbrace
%\end{equation}

%\begin{frontmatter}

\title{A Revision of a New Chaos-Based Image Encryption System: Weaknesses and Limitations }

%% \author[label1]{Hassan Noura}\ead{hnouran@gmail.com}
%% \author[label2]{Lama Sleem}
%% \author[label2]{Raphael Couturier}
%% \address[label1]{Lebanese University, Faculty of Engineering, Department of Computer Science and Telecommunication, Beyrouth, Lebanon.}
%% \address[label2]{FEMTO-ST Institute, Univ. Bourgogne Franche-Comt\'e (UBFC), France}

\author{
\IEEEauthorblockN{Hassan Noura\IEEEauthorrefmark{1}, Lama
  Sleem\IEEEauthorrefmark{2}, Raphaël Couturier\IEEEauthorrefmark{3}}
\IEEEauthorblockA{\IEEEauthorrefmark{1}Lebanese University, Faculty of Engineering,
  Department of Computer Science and Telecommunication, Beyrouth,
  Lebanon\\
  \IEEEauthorrefmark{2}\IEEEauthorrefmark{3}FEMTO-ST Institute, UMR 6174 CNRS - Univ. Bourgogne Franche-Comt\'e (UBFC), Belfort, France\\
Email:
\IEEEauthorrefmark{1}hnouran@gmail.com,\IEEEauthorrefmark{2}lama.sleem@univ-fcomte.fr,\IEEEauthorrefmark{3}
raphael.couturier@univ-fcomte.fr
}

}

\maketitle

\begin{abstract}

%%%%%%%%%%%
Lately, multimedia encryption has been the focus of attention in many researches. Recently, a large number of encryption algorithms has been
presented to protect image contents.The main objective of modern image
encryption schemes is to reduce the computation complexity in order to
respond to the real time multimedia and/or limited resources
requirements without degrading the high level of security. In fact,
most of the recent solutions are based on the chaotic theory. However,
the majority of chaotic systems suffers from different limitations and
their implementation is difficult at the hardware level because of the non
integer operations that are employed requiring huge resources and
latency. In this paper, we analyze the new chaos-based image encryption system
presented in~\cite{el2016new}.
It uses a static binary diffusion layer, followed by a key dependent
bit-permutation layer that only iterates for one round.
Based on their results in this paper,  we claim that the uniformity
and avalanche effect can be reached from the first round. However, we tried to
verify the results but our conclusion was that these results were wrong
because it was shown that at least 6 iterations are necessary to ensure the required cryptographic performance such as the plain-sensitivity property. Therefore, the required execution time must be multiplied
by 6 and consequently this will increase the latency.
In addition to all aforementioned problems, we find that ensuring the avalanche effect in the whole
image introduces a high error propagation. In order to solve this problem, we recommend to ensure the avalanche effect in the level of blocks instead of the whole image.
\end{abstract}


%\end{frontmatter}
% Additional space between abstract & rest of document

%\begin{multicols}{2}
% \linenumbers

\section{Introduction}
%\RC{abstract too long... I tried to comment non necessary sentences}
%\HN{yes.}
Encrypting images is a necessary requirement to protect the privacy of people and the confidentiality of image contents. However, traditional cryptographic techniques, using symmetric-key standard encryption algorithms such as DES~\cite{biham1993differential} and AES~\cite{daemen2002design}, are not efficient for encrypting images and video contents due to the intrinsic features of images, and the strong correlation among the adjacent pixels~\cite{flayh2009wavelet}. Therefore, traditional encryption techniques cannot fit the real-time delivery of multimedia streams according to~\cite{dan2008image} and discovering another solution becomes absolutely necessary.\\

Another paradigm has been investigated by researchers in the last decade, which is the "Chaos" field. Chaos consists of a non-linear dynamic system that appears to be random. Due to the extreme sensitivity to initial conditions, chaos was integrated extensively to build the cryptographic algorithms of digital images such as in \cite{seyedzadeh2012fast,tong2009new, akhshani2010novel, seyedzade2010novel,kumar2011extended,zhu2011chaos,huang2013implementation, borujeni2013chaotic}. Unfortunately, chaos-based encryption algorithms are not always secure, and most of them have been successfully crypt-analyzed~\cite{li2011breaking,rhouma2010cryptanalysis, li2002cryptanalysis}, due to their instability coming from the periodicity of mapping~\cite{huang2009security} and the finite computing precision that renders the system vulnerable to different kinds of attacks~\cite{arroyo2009cryptanalysis,alvarez2009cryptanalyzing}. Additionally, the main disadvantage of the majority of chaotic encryption algorithms is the use of \textbf{floating calculations} which makes the practical software or hardware implementation of such systems not efficient and  complex compared to the traditional ciphers such as AES and DES, which only operate with integer operations.\\

Recently, a scheme was presented in~\cite{el2016new}. It consists in
applying a static binary diffusion layer followed by a key dependent
bit-permutation based on the periodicity 2D cat map that only iterates 
for \textbf{one round}. In fact, the authors indicated that the avalanche
effect can be reached from the first round.

\subsection{Motivation \& Contributions}
\label{subsec:Motivation}
The main motivation of our work is to analyze the previous work
of~\cite{el2016new} and to quantify its security and performance level to validate whether it can be a good encryption candidate or not.

%\RC{reread the next sentence because I changed it}
%\HN{done}

In this paper, we detail all the weaknesses of this proposal and we
prove that the avalanche effect is not attained. In addition, we focus on the weak key size
permutation technique and its inflexibility that \textbf{limits} the use of this cipher.

However, according to the tests done in this paper, it is shown that
the sufficient number of iterations $r$ needed to reach the avalanche
effect (plain-sensitivity property) is $\geq6$. Therefore,
the required execution time of this scheme should be multiplied by
6. Based on the obtained results, we demonstrate that~\cite{el2016new} does not meet the main contribution needed which is a lower latency. Therefore, we prove that the required latency and resources are so high to ensure the required robustness against attacks. In fact, with $r=1$, the system is considered weak against different kinds of attacks such as known/chosen plain-text/cipher-text attacks. In addition, it cannot be used by tiny devices such as our smart-phones or sensors because of the high required memory size. More important, we validate that it is extremely sensitive against the channel error thus preventing its practical implementation. We highlight all these points and give the correct results that should have been taken into consideration.
\subsection{Organization}

The rest of this paper is organized as follows. In Section
\ref{sec:related}, an analysis of the proposed scheme of~\cite{el2016new} is presented and
the weak points are detailed. Moving to Section ~\ref{sec:Algo},
the false result of the avalanche effect is proved in addition to quantify the difference between original and cipher images.. Furthermore, the uniformity
analysis is done. Then, in Section~\ref{sec:security}, the propagation
of error analysis shows that any noise affecting the channel will
prevent the recovery of the original contents after decryption. The visual
degradation analysis was done by studying PSNR~\cite{huynh2008scope}
and SSIM~\cite{wang2004image}. Then, a general performance analysis in
addition to the execution time is discussed in Section~\ref{complexity}. Finally, in Section~\ref{sec:conclusion}, a conclusion summarizes our work and future works are presented.

\section{Analysis of the proposed chaos-based image encryption system }
\label{sec:related}
In this section, we analyze the cipher scheme of~\cite{el2016new} that
is illustrated in~\figurename~\ref{figemmiter}. In fact, several points are presented and prove that the proposed cipher cannot reach the efficiency and the required cryptographic performance. As a result, this solution can be seen as unsafe. All details are shown in the following:

\begin{figure*}[!ht]
\begin{minipage}{\linewidth} 
\centering
\subfloat[][Encryption Algorithm]{\includegraphics[scale=0.41]{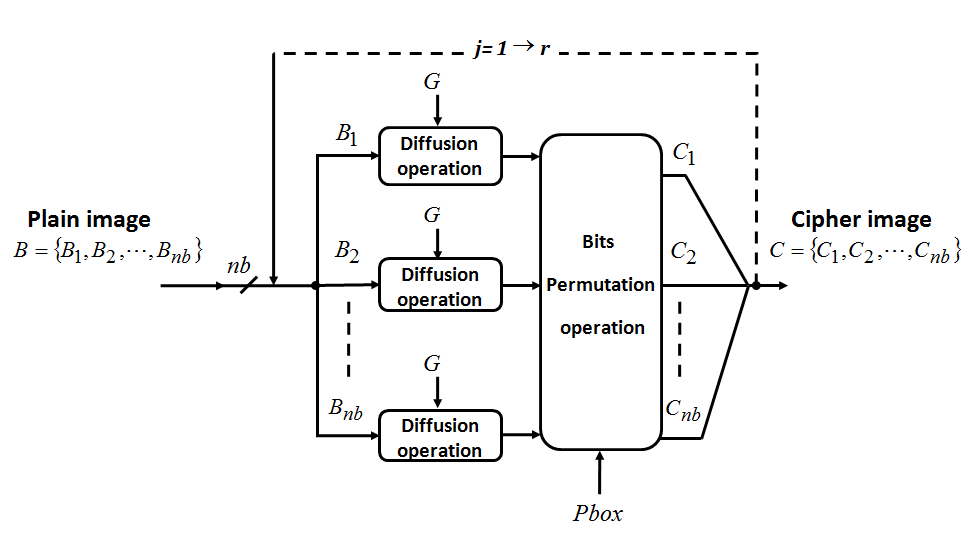}}
\subfloat[][Decryption Algorithm]{\includegraphics[scale=0.38]{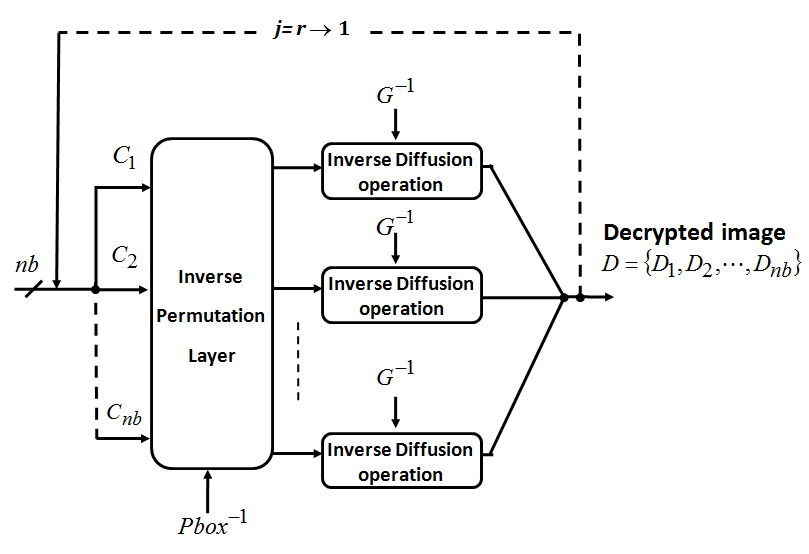}}
\caption{Architecture of the encryption(a) and decryption (b) algorithm described ~\cite{el2016new}.}
\label{figemmiter}
\end{minipage}
\end{figure*}

%\RC{what is rp, M?}
%\HN{Verifiy text}
\begin{enumerate}

\item {\textbf {Low cryptographic performance and attacks
    vulnerability }}: A low number of rounds $r$ based on wrong
  results enables to reduce the execution time. More important, the
  security level here is considered low since the plain sensitivity is
  very low. This makes chosen/known plain-text attacks easy. Additionally, the size of the permutation key used is $4\times q$, where $q=\lceil{log_2(M)}\rceil$ is dependent on the size of the square input image ($M\times M$), which is feasible for any brute force attack and specifically for small sized images. Therefore, this algorithm cannot resist different kinds of attacks. Also, the static structure of the binary diffusion matrix which is independent of the key is not preferable from cryptographic viewpoints  and should be dependent on a generated key to strengthen its security level. Also, the dimension of the diffusion matrix is fixed, which removes the flexibility property.\\

Moreover, in their paper, the authors considered that the permutation was done only once. 
Therefore, the chaotic generator should be iterated once to produce
the required parameters and initial conditions of the permutation
technique that has 32 bit length for $M\leq 256$. In fact, this is
not sufficient to ensure a
sufficient length of permutation sub-key and consequently cannot resist the brute force attacks and is not enough to reach the required cryptographic performance~\cite{paar2009understanding} (see chapter 3).
%\RC{Ref for the previous statement?}
%\HN{Done}

On the other hand, the generator may require to iterate twice for $M\geq 256$ that means that the length of the required static pseudo-random bits is 64 and it is also not sufficient to break the brute force attacks~\cite{paar2009understanding}. 
%%%%%%%%%%%%%%%%%%
% lama verify it please... recent
%%%%%%%%%%%%%%%%%
In addition to that, as the round of permutation $rp$=1 and the permutation technique is periodic, this means that the produced permutation table has a short period according to~\cite{wang2009chaos}. This paper \textbf{ validates the employment of different parameters and initial conditions for several iterations to prevent the periodicity of permutation}. 

%\RC{define page}
%\HN{done}
\item \textbf{Not flexible}: The employed permutation technique is
  based on the Noura formulation of the 2D cat map that was previously presented in~\cite{nouratel01104996}(see page 90-91). Additionally,
  the proposition of using invertible binary diffusion matrix for
  image encryption was also presented previously
  in~\cite{nouratel01104996}(see page 113-121) and it is not the
  original proposition of~\cite{el2016new}. Unfortunately, this
  permutation technique is not flexible and it requires the size of
  the original image to be square. This is another limitation for this approach since a good approach must be flexible for any dimension desired by the user.
  
 \item \textbf{High size of memory}: Employing  bit permutation instead of byte permutation increases the required memory size by a factor of 8, which is unacceptable for different systems. For example, tiny devices are not able to apply this approach since their memory is limited.

\item {\textbf {Higher latency}}: The design of the round function
  with the goal to reach the avalanche effect with lower round number is not achieved. Moreover, the bit
  permutation algorithm is realized in the bit level and not in the
  byte level as in AES. This introduces an overhead in terms of execution time since the operation in byte or word level as in AES is more efficient compared to the bit level as in DES.
\item {\textbf {Difficulty to adapt to modern devices:}} Low execution time and low memory requirements are all mandatory conditions to have an efficient cipher that can take into consideration the short life of batteries and the low resources available especially for tiny devices such as sensors or smart phones.
\item {\textbf{High error propagation}}:  The analysis scheme has a trade-off between the avalanche effect and the propagation error. In fact, the obtained result indicates that a random bit error can destroy the contents of the image, which is not suitable for some applications, especially in wireless communication where the channels are subjected to different kinds of noises~\cite{alajel2010error}.
\end{enumerate}
Therefore, all these challenges and weaknesses clearly indicate that the cipher scheme of~\cite{el2016new} is neither efficient nor secure and cannot be considered as a good image encryption candidate. 
\section{Analysis of the Avalanche effect, Sensitivity, and Uniformity}
\label{sec:Algo}
\subsection{Avalanche effect}
Indeed, authors of~\cite{el2016new} indicated that the proposed cipher can reach the Avalanche effect after one iteration ( $r$=1 see \figurename~7
in~\cite{el2016new}), which is not true and it was obtained for a small image size $16\times
16$. Moreover, authors of~\cite{el2016new} generalized the result to one round for the different sizes of any image, which is not logic. Therefore, the
avalanche effect test (see the next subsection by applying the
sensitivity test) was applied on the described scheme and a
different result  was found that is shown clearly in Table~\ref{obtainedresultsavalanche} for different sizes of images. It is clear, in this table, that 6 iterations are needed to reach an avalanche effect near 50\%.
%%%%%%%%%%%% lama verify it
 \subsection{Sensitivity}
The sensitivity refers to a huge change in the cipher text in response to a slight change in the original message itself. A cipher algorithm$E_K()$ is considered to be robust against chosen/known plain-text attacks, if it ensures the avalanche effect. In other words, the percentage of the Hamming distance (in bits) between the corresponding cipher image $C=E_k(I)$ and $C'=E_K(I')$ should be close to 50\%, while $I$ and $I'$ only differ by one bit. An image $I$ has been chosen that have all values equal to zero. $I'$ is different from $I$ by only one random pixel that has value equals to one.
The sensitivity of the plain image $PS$ is analyzed for 1000 random plain images and keys using the percentage of the Hamming distance,which is calculated as follows:

%\RC{$E_{K_w}$ needs to be defined}
%\HN{done}

\setlength{\arraycolsep}{0.0em}
\begin{eqnarray}
PS_w=&{}{}&\frac{\sum (Byte2bit (C_w)  \oplus Byte2bit(C_w') ) }{T}\times 100\%\\ \nonumber\\
&& =\frac{\sum Byte2bit(E_{K_w}(I)) \oplus Byte2bit(E_{K_w}(I'))}{T}\times 100\%                       \nonumber
\label{eqn_aval}
\end{eqnarray}
where $T$ is the length in bits of the original and cipher images, $C_w$ and $ C_w'$ are the corresponding $w^{th}$ cipher images using $I_w$ and $I_w'$ and ($K_w$) secret keys respectively. All the elements of $I_w'$ are equal to those of $I_w$, except a random Least Significant Bit ($LSB$) of a random byte, which was flipped and $w=1,\;2,\;\ldots , \;1000$.

  \begin{figure*}[!ht]
\centering
\begin{minipage}{\linewidth} 
\centering
\subfloat[][$PS$]{\includegraphics[scale=0.4]{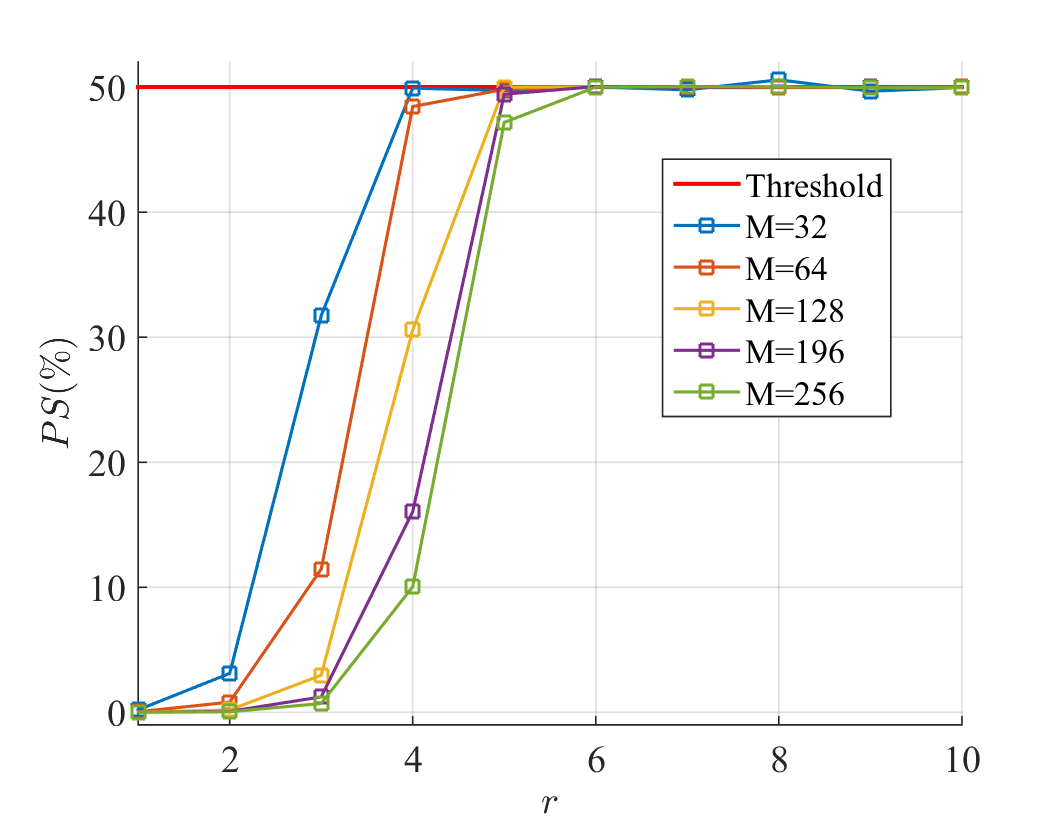}}
\subfloat[][$Diff$]{\includegraphics[scale=0.4]{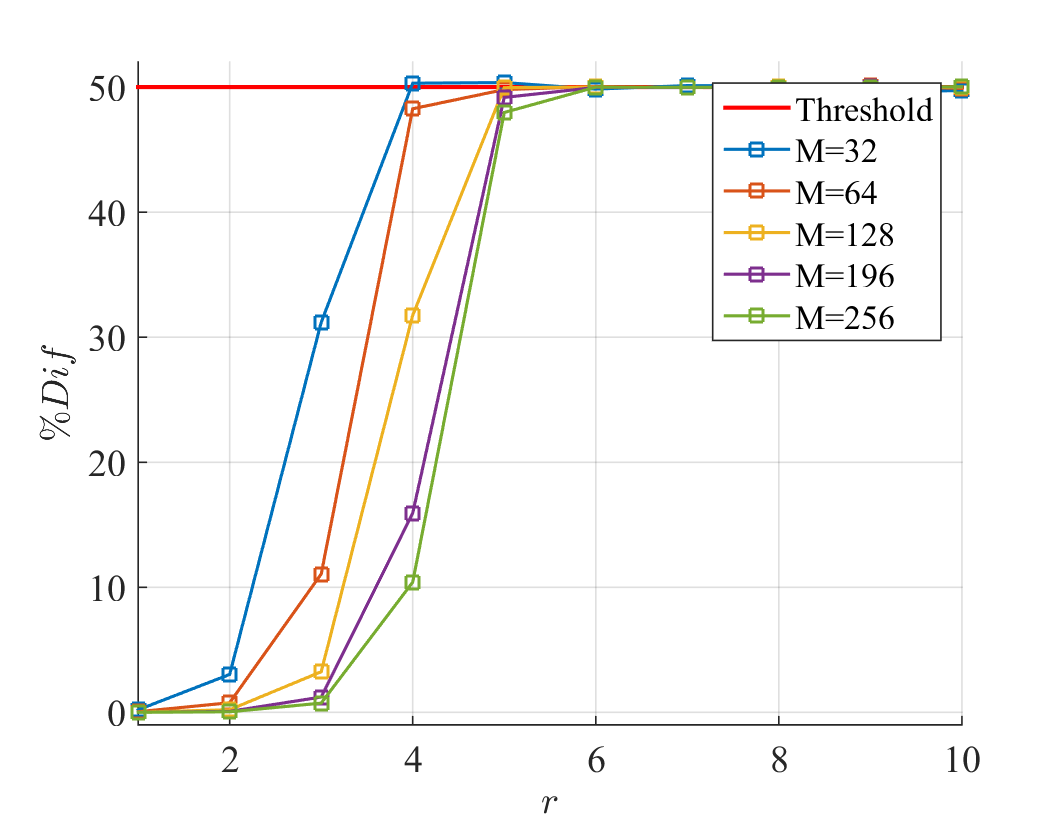}}
\caption{Variation of the average of the percent of the avalanche effect and the mean difference between plain and cipher-text over 1000 random keys versus the number of rounds $r$ respectively.}
\label{fig:PS}
\end{minipage}
\end{figure*}
  
According to the obtained results, the sufficient number of rounds, $r$ should be $\geq$6 to be dependent on the size of images as shown in Table \ref{obtainedresultsavalanche} and Figure \ref{fig:PS}, where the size of images is square ($M \times M$). Therefore, to reduce the execution time with a good avalanche degree, $r$ must be set to 6 for $M<196$ and $r>6$ for $M>=196$. Finally, $r$ equals to 6 is sufficient to reach the avalanche effect for the different analyzed sizes of images (the maximum size of simulation that was analyzed was $512\times 512$). Moreover, the obtained results illustrated here are reasonable, since increasing the image size will definitely increase the number of blocks, which requires a sufficient number of rounds to propagate the difference among blocks to reach the avalanche effect. Based on that, $r$ cannot be set to 1 as indicated in~\cite{el2016new}. Therefore, a lower degree of avalanche effect is reached, which means that it cannot immune the system against chosen/known plain-text attacks according to their configuration($r=rp=1$). Therefore, this cipher is insecure with this number of rounds. Consequently, after these results, $r$ must be 6 to obtain the required level of security. On the other hand, this will lead to an increase in the execution time.

\begin{table*}[!t]
\centering
\caption{Variation of the avalanche effect versus the size of image and the number of rounds $r$.}
\begin{tabular}{|c|c|c|c|c|c|c|c|}
\hline
\textbf{\backslashbox{$M$}{$r$}} & \textbf{1} & \textbf{2} & \textbf{3} & \textbf{4} & \textbf{5} & \textbf{6} & \textbf{7} \\ \hline
\textit{$16\times16$}  &0.3092  &	4.2415  &	29.052  &	45.09	  &49.39 &	\textbf{50.03}  &	49.97           \\ \hline
\textit{$32\times32$} &0.077  &	1.135	  &11.54  &	40.26	  &48.29	  &\textbf{49.6}	  &49.86          \\ \hline
\textit{$64\times64$}  &0,0193  &	0,342  &	5,720  &	44,70	  &49,99	  &\textbf{50,08}	  &49,64          \\ \hline
\textit{$128\times128$} &0.004	  &0.082  &	1.407  &	15.00	  &47.34	  &\textbf{50.03}	  &49.9          \\ \hline
\textit{$196\times196$} &0.0021   &   	0.035    &  	0.52      &	4.51   	  &39.80     &	\textbf{49.95}     &	49.99       \\ \hline
\textit{$256\times256$} &0.0012  &	0.0212  &	0.3366	  &5.348	  &42.441  &	\textbf{50.01}	  &49.97         \\ \hline
\textit{$300\times300$} &0.0009	  &0.0152  &	0.2588	  &4.290	  &38.273	  &\textbf{50.04}	  &50.12          \\ \hline
\textit{$512\times 512$} & 0.0003	  &0.005	  &0.087  &	1.429	  &19.688  &	\textbf{49.99	}  &50.02         \\ \hline
\end{tabular}
\label{obtainedresultsavalanche}
\end{table*}

Figure~\ref{fig:PS}(a) shows the average values of the percentage of the Hamming distance over 1000 random dynamic keys versus the number of rounds $r$. The obtained results show that the optimal number of iterations to reach the avalanche effect is $6$ for the different sizes of images.
Additionally, to ensure the independence between the plain and cipher images, the percentage of the Hamming distance between $I$ and its corresponding encrypted one $C$ are computed as described above (see Eq.~\ref{eqn_aval}). The obtained results in Figure~\ref{fig:PS}-(b) show that the independence between the plain and cipher images requires the same number of iterations of the avalanche effect. This means that to reach the independence and the avalanche effect, $r$ should be $\geq$ 6.

\subsection{Analysis of the uniformity}
To resist the common statistical attacks, the encrypted image should possess certain random properties. The most important one is that the frequency of each symbol of the encrypted image should be uniform. This means that each symbol has an occurrence probability close to $\frac{1}{n}$, where $n$ is the number of symbols. In order to compute the level of uniformity of each encrypted image, the \textbf{Chi-square test} is applied as expressed in Equation~\ref{eqp13}:
\vspace{-0.3cm}
\begin{equation} 
\chi^2= \sum_{i=0}^{Q-1}\frac{(o_i - e)^2}{e}
\label{eqp13}
\end{equation}

\begin{figure*}[!ht]
\centering
\begin{minipage}{\linewidth} 
\centering
\includegraphics[scale=0.4]{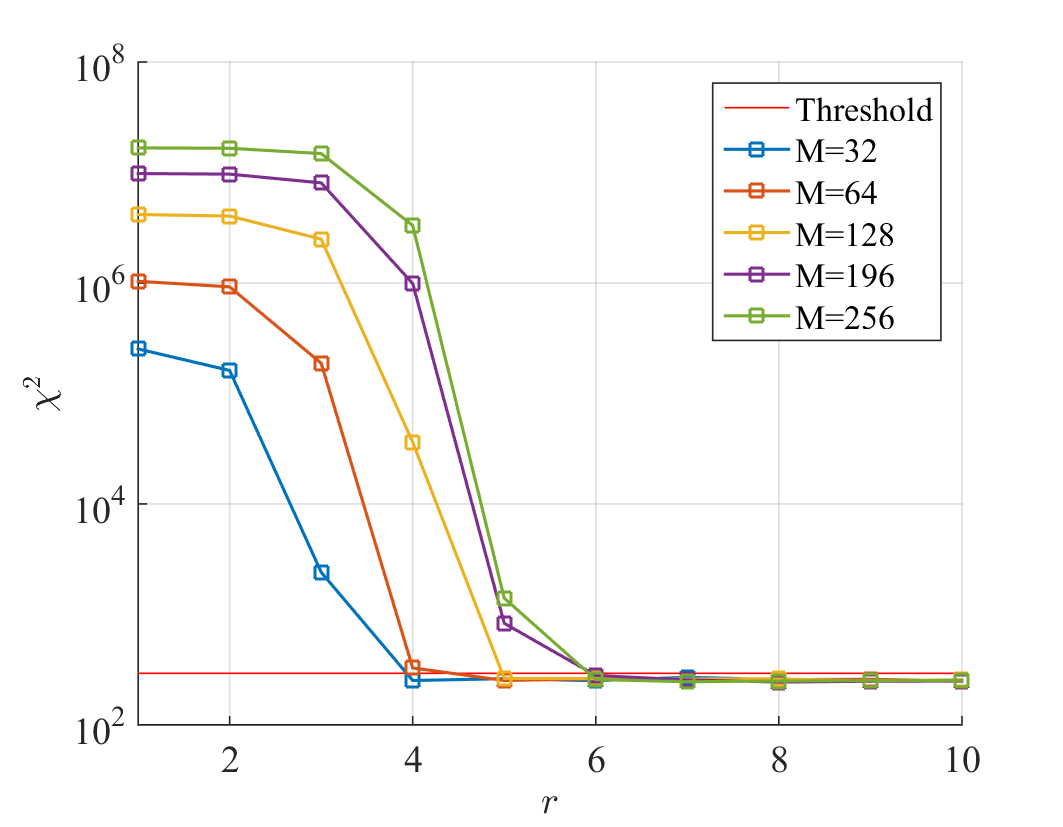}
\caption{Variation of the average of the chi-square test over 1000 random keys versus the number of rounds $r$ respectively, with $Tb=256$.}
\label{fig:unif}
\end{minipage}
\end{figure*}

Where $Q$ is the number of gray levels (here we work with gray scale images, Q=256), and $o_i$ is the observed occurrence frequencies of each gray level (0-255) and $e$ is the desired uniform frequency that equals to $\frac{len}{Q}$, where $len$ is the length of image in byte level.
This statistical test is used to compare the observed data with a specific hypothesis. Hence, the null hypothesis is formulated, which is then rejected or retained with the help of some statistical tests. The probability value below, which is the null hypothesis, is rejected and is called the alpha level or simply the "significant level". It is conventional to conclude that the null hypothesis is false if the probability value is less than 0.05 \cite{du2009confidence}. The level of significance of 0.05 (or 5$\%$) is often chosen. In fact, with a significance level of 0.05, researchers can be 95 $\%$ confident that the results represent a non-chance finding ~\cite{vanvoorhis2007understanding}. Indeed, with a significant level of $0.05$ and a number of intervals equal to 256, the chi-square reaches a maximal value equals to $293$ ~\cite{chen2015efficient}. So, all values lower than this value are acceptable and indicate the uniformity distribution of the histogram.
This criterion is verified by testing the chi-square for the previous mentioned images $I$ under 1000 different dynamic keys. Figure \ref{fig:unif} shows that the mean chi-square values become $\leq 293$ after 6 iterations for all the dimension. This confirms the previous result and clearly indicates that $r\geq 6$ is sufficient to reach the independence avalanche effect, and uniformity property of the encrypted image under the proposed algorithm.

However, ensuring the security by reaching the avalanche effect introduces a hard challenge that is described in the following. This will have a high impact of a single bit change in the encrypted image and consequently a hard visual degradation is obtained and these results are described in the next section. This indicates clearly that a trade-off between the avalanche effect and error propagation is reached with this  kind of scheme.
\section{Propagation of errors}
\label{sec:security}
\begin{figure*}[!ht]
\centering
\includegraphics[scale=0.5]{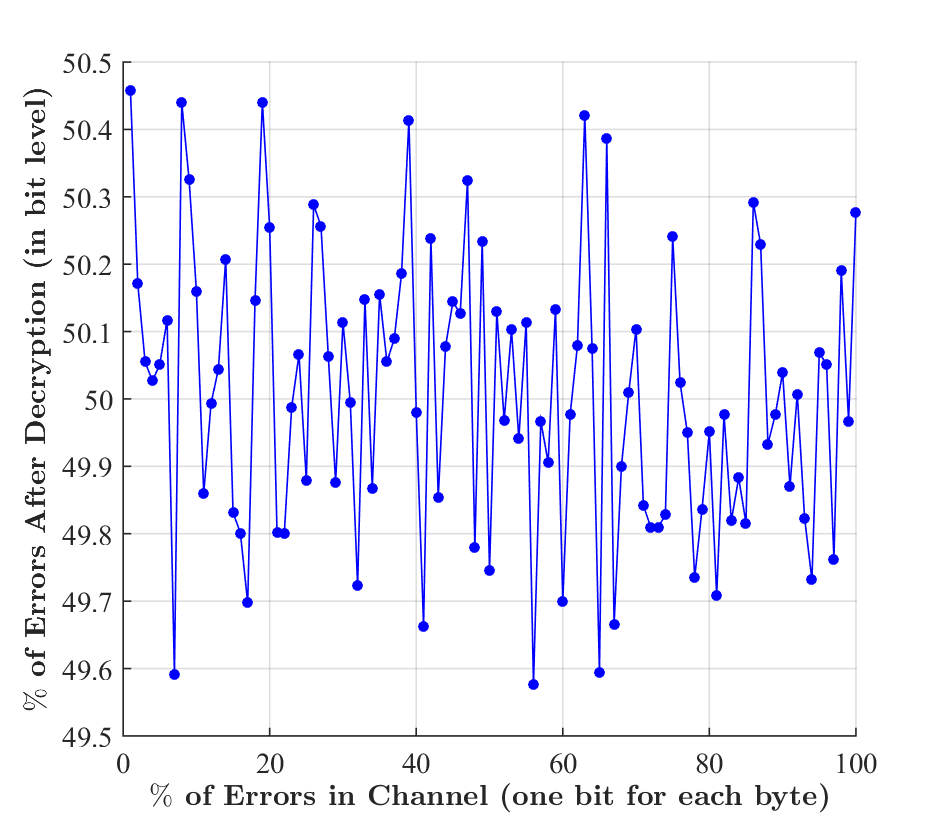}
\caption{Variation of the impact of the error propagation according to the percentage of errors.}
\label{fig:error}
\end{figure*}

Indeed, an important criterion that should be ensured for any cipher is
the tolerance error, which means that the error is not
propagated. Interference and noise existing in the transmission
channel are the main causes of error. However, a bit
error means that a substitution of '0' bit into '1' bit or vice versa
will take place. This error may propagate and leads to the destruction
of data, which is a big challenge since a trade-off between the Avalanche
effect and an error propagation is shown in ~\cite{massoudi2008overview}.
  In~\cite{el2016new}, if a bit error takes place in any encrypted block, it will affect the overall decrypted image and the difference between both decrypted images are calculated according to Equation~\ref{eqn_aval}. The result is presented in Figure~\ref{fig:error} which shows that the error is always close to 50\%.
  As a result, we can deduce that the proposed approach is inefficient
  to overcome the propagation error. Therefore, any error in one byte will propagate to all other bytes which makes the system powerless against noisy and fading channels.

\subsection{\textbf{Visual Degradation}}

This test is specific for image and video contents and enables to quantify the visual degradation that it reaches by employing the cipher scheme with the effect of error propagation. In fact, the degradation operated on the decrypted image after any single change of any bit prevents the contents of an image from being recognized. To measure the visual degradation, two well known parameters are studied to measure the encryption visual quality which are Peak Signal-to-Noise Ratio (PSNR) and Structural Similarity Index (SSIM). 

\begin{figure*}[!ht]
\begin{minipage}{\linewidth}
\begin{center}
\subfloat[][]{\includegraphics[scale=0.5]{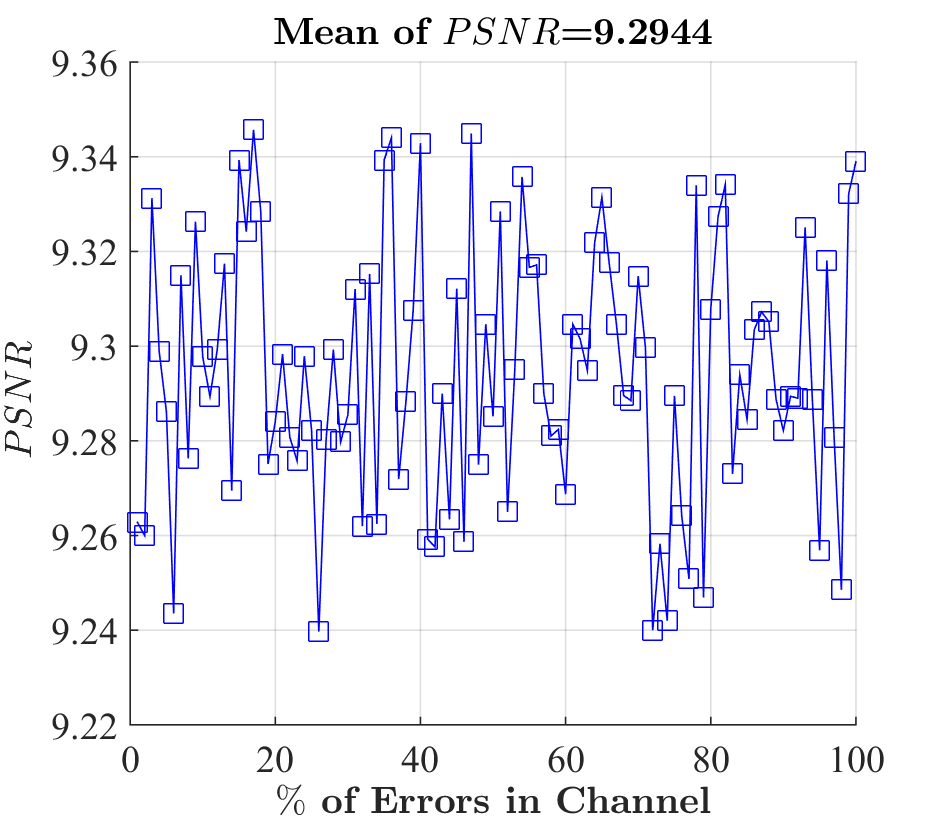}}
\subfloat[][]{\includegraphics[scale=0.5]{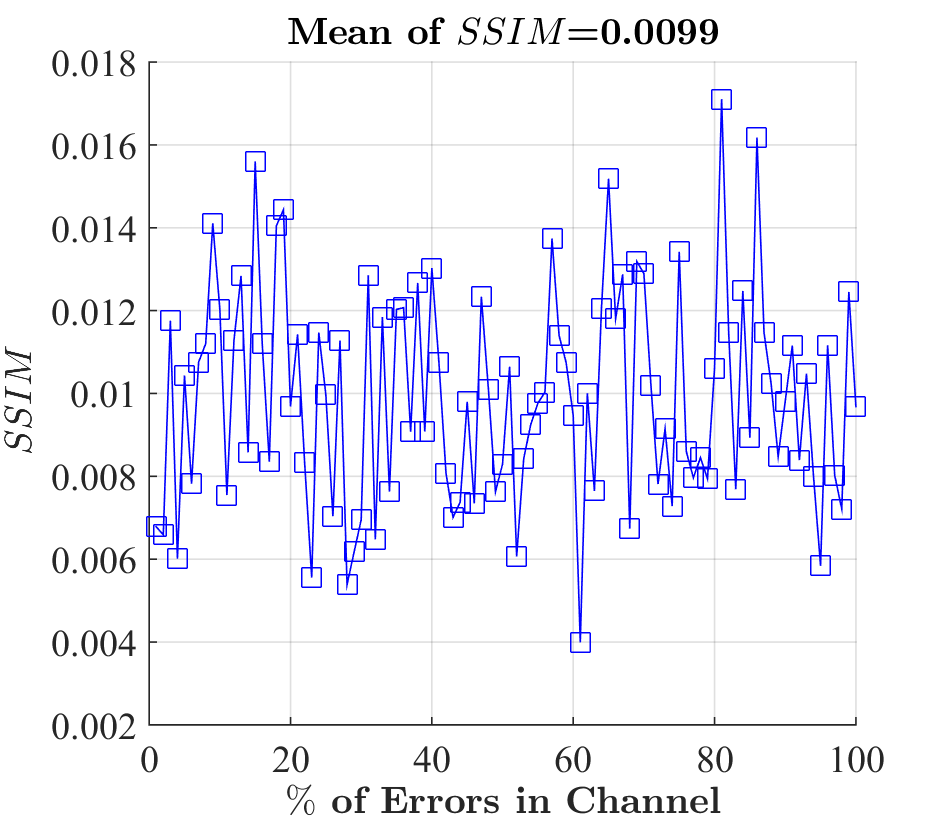}}
\end{center}
\vspace{-0.5cm}
\caption{$SSIM$ variation between both decrypted Lenna images versus $1000$ dynamic keys versus the percentage of errors in channel.}
\label{fig:PSNR and MSSIM}
\end{minipage}
\end{figure*}

PSNR is derived from the Mean Squared Error (MSE), while MSE represents the cumulative squared error between an original and an encrypted image. A lower PSNR value indicates that there is a high difference between the original and the cipher images. \\
%\RC{next sentence is not clear}
%\HN{done but verify}
In addition, another metric is used and called Structural SIMilarity (SSIM) index~\cite{li2002cryptanalysis}, which is defined after the Human Visual System (HVS) and quantifies the similarity between two images. SSIM is in the interval [0,1]. A value of 0 means that there is no correlation between the original and the cipher images, while a value close to $1$ means that both images are approximately the same. In this context, $PSNR$ and $SSIM$ are measured  between two decrypted Lena images (where the  second decrypted image  corresponds to the encrypted image with a percent of error). Indeed, in ~\figurename~\ref{fig:PSNR and MSSIM}-(a) and (b), PSNR and SSIM are shown respectively. As shown, the variation of PSNR  and SSIM versus the percentage of errors is presented. This low value validates that the proposed cipher provides a high difference between both decrypted images from the lower error percentage.  This means that a high and hard visual distortion is obtained from a small error percentage. Therefore, the cipher algorithm cannot be employed in practical systems that suffer from the error channel.

Moreover, in Table~\ref{table:statist} the results are clearly showing the values of PSNR, SSIM and $Dif$. Dif is the sensitivity test that measures the difference between two decrypted images with a percentage of errors introduced in the encrypted image (we suppose that it is the effect of a channel noise in the encrypted image). It is close to 50\%, which is relatively a high value that will prevent the recovery of the original image.
As a conclusion, the proposed scheme suffers from the hard visual degradation caused by the channel error. This means that no useful visual information or structure about the original image could be revealed from the decrypted image if any error in the channel is introduced.

\begin{table*}[!ht]

\centering
\caption{Statistical results of sensitivity for Lenna image for 1000 random keys.}
\begin{tabular}{ll|l|l|l|l|}
\hline
\multicolumn{6}{|l|}{$\;\;\;\;\;\;\;\;\;\;\;\;\;\;\;$ Integer Diffusion matrices $\;\;\;\;\;\;\;\;\;\;\;\;$ }                 \\ \hline
\multicolumn{2}{l|}{}   & Min  &  Mean &  Max & Std  \\ \hline
\multicolumn{2}{|l|}{$Dif$} &  49.7505   &50.0091   &50.2399   & 0.1014\\\hline
\multicolumn{2}{|l|}{PSNR} &   9.2397  &   9.2944 &    9.3457 &    0.0275\\   \hline
\multicolumn{2}{|l|}{SSIM} & 0.0040&    0.0099 &   0.0171  &    0.0026 \\ \hline
\end{tabular}
\label{table:statist}
\end{table*}

\section{\textbf{Performance Analysis}}
\label{complexity}
In practice, it is very  important that the cipher requires less
latency, memory and lower resources for the ciphering/deciphering process
in order to be considered efficient.
The presented encryption scheme employs bit permutation on the overall image, which requires a huge memory size and prevents its employment in tiny devices. The proposed scheme must undertake a higher number of rounds ($r=6$) and consequently requires 5 times overhead in addition to the one presented in~\cite{el2016new}. Thus, this proposal is not efficient in tiny devices which are battery limited and cannot respond to the needs of real time applications. In addition, the required size of memory and the execution time should always be at a minimum to reach a good encryption candidate.   

\subsection{Discussion and Cryptanalysis: Resistance against the well-known types of attacks}
In the following, the typical cryptanalytic cases appearing in the literature are considered and a brief analysis of the proposed cipher against several cryptanalytic attacks is provided. The proposed cipher algorithm is considered to be public and the cryptanalyst has a complete knowledge about the employed confusion and diffusion primitives to know the technique used to build them but no knowledge about the secret key is available. The previous scheme of~\cite{el2016new} can be broken by employing different types of attacks. It is not sufficient to achieve the required level of security since the scheme fails to resist statistical attacks (uniformity is not attained) in addition to chosen/known plain-text attacks. Therefore, differential attacks are based on studying the relation between two encrypted images resulting from a slight change, usually one bit difference compared to the original one. A successful sensitivity test shows how much a slight change in the plain-image or in the key will affect the resulted cipher image. 
Moreover, the key space used is fixed and can be $2^ {4\times q}$, which is not sufficient to prevent the brute-force attack.
  Finally, the problem of single image failure and accidental key disclosure is not taken into consideration by this scheme. Furthermore, differential and linear attacks would become effective. Therefore, this cipher cannot resist different kinds of attacks.
\section {Conclusion and Future Work}
\label{sec:conclusion}
\vspace{-0.2 cm}
%\RC{The end of the first sentence is not clear for me}
%\HN{done}

In this paper, we analyzed the previous cipher presented in~\cite{el2016new}. In fact,  we proved that this scheme has different weaknesses such as the wrong avalanche effect reached in addition to a short size of permutation key. This leads to consider the system insecure against different kinds of attacks. More important, error propagation was studied and its results were devastating on the decryption side since no useful data can be obtained when any bit is subjected to a channel noise.
 Therefore, this solution is not efficient since it requires a huge memory size, \textbf{higher} latency and cannot resist the channel error. In addition, it is not secure since it cannot face powerful attacks. These results are presented in order to prove the non credibility and the unsafe employment of~\cite{el2016new}. For future works, we aim to build an efficient and secure lightweight image encryption scheme that will overcome all the stated challenges details in this paper.

\bibliographystyle{IEEEtran}
\bibliography{bibfile}

%\end{multicols}
\end{document}